\documentclass[12pt]{article}
\usepackage{amsmath,amsthm,amsfonts}

\newcommand{\NN}{{\mathbb N}}
\newcommand{\RR}{{\mathbb R}}

\newcommand{\HH}{{\mathbb H}}
\newcommand{\dd}{\mathrm d}

\newcommand{\pd}{\partial}

\newcounter{mylc}
\renewcommand{\themylc}{\roman{mylc}}

\newtheorem{theorem}{Theorem}
\newtheorem{proposition}{Proposition}

\theoremstyle{remark}
\newtheorem{remark}{Remark}
\newtheorem{example}{Example}

\theoremstyle{definition}

\begin{document}
\title{A geometric approach to the equilibrium shapes of self-gravitating fluids}

\author{D. Peralta-Salas\thanks{dperalta@fis.ucm.es}}
\date{\normalsize Departamento de F\'{\i}sica Te\'orica II, Facultad de
Ciencias F\'{\i}sicas, Universidad Complutense, 28040 Madrid,
Spain}
\maketitle
\begin{abstract}
The classification of the possible equilibrium shapes that a
self-gravitating fluid can take in a Riemannian manifold is a
classical problem in mathematical physics. In this paper it is
proved that the equilibrium shapes are isoparametric submanifolds.
Some geometric properties of the equilibrium shapes are also
obtained, specifically the relationship with the isoperimetric
problem and the group of isometries of the manifold. This work
follows the new geometric approach to the problem developed by the
author.
\end{abstract}
\section{Introduction}
Let $(M,g)$ be an analytic, complete and connected $n$-dimensional
Riemannian manifold and $\Omega$ an open connected subset of $M$
(bounded or not) occupied by a mass of fluid. We say that a fluid
is self-gravitating if the only significative forces are its
interior pressure and its own gravitation. Depending on whether
the gravitational field is modelled by Poisson or Einstein
equations we say that the fluid is Newtonian or relativistic. An
important problem in Fluid Mechanics consists in studying the
shape that a self-gravitating fluid will take when it reaches the
equilibrium state. By the term shape of a fluid I mean the
topological and geometrical properties of the boundary $\pd
\Omega$. The mathematical description of this kind of fluids only
involves three physical quantities, the gravitational potential,
which is a function $f_1:M \to \RR$, and the density and pressure,
which are two analytic ($C^\omega$) functions
$f_2,f_3:\Omega\to\RR$. The set of partial differential equations
that the functions $(f_1,f_2,f_3)$ must verify is of free-boundary
type because the domain $\Omega$ is an unknown of the problem. In
the relativistic case the metric tensor $g$ is also an unknown and
it must satisfy the coupling condition
$R_{ab}=f_1^{-1}f_{1;ab}+4\pi(f_2-f_3)g_{ab}$, $R_{ab}$ standing
for the Ricci tensor and the semicolon standing for covariant
derivative.

The standard approaches to the problem of classifying the
equilibrium shapes of $\pd \Omega$ generally employ analytical
techniques. In the Newtonian case maximum principles for elliptic
equations are used in order to prove the existence of symmetries
of the solutions $(f_1,f_2,f_3)$. Lichtenstein \cite{Lich} and
later on Lindblom \cite{Lind} proved the existence of spherical
symmetry (i.e. $\pd \Omega$ is a round sphere) when $M$ is the
Euclidean 3-dimensional space, $\Omega$ is bounded and the
functions $(f_1,f_2,f_3)$ satisfy some physical constraints. In
the relativistic case arguments involving the positive mass
theorem are used for obtaining the conformal flatness of the
metric tensor. As a consequence of this technique Beig\&Simon
\cite{BeSi} and Lindblom\&Masood-ul-Alam \cite{LiMa} proved again
spherical symmetry when $\Omega$ is bounded and the solutions
verify hard physical constraints. Despite of these important
results many questions remain open: What about Newtonian fluids in
Riemannian manifolds? What about unbounded domains $\Omega$? Do
the same results hold if we drop the physical constraints?

In this work we follow the geometrical theory for studying
equilibrium shapes developed by the author in \cite{PP-S}. The
system of equations that we consider is the following, which we
set up as problem $(P1)$

\begin{gather}\label{P1}
 \Delta f_1 = F(f_1,f_2,f_3)\,\, \text{in}\,\, \Omega \\
 H(f_1)\nabla f_3 + G(f_2,f_3)\nabla f_1=0\,\, \text{in}\,\,\Omega\\
 f_1=c,\, c \in \RR,\, \nabla f_1 \neq 0 \,\,\text{and}\,\, f_1 \in
 C^2_t\,\, \text{on}\,\,
 \pd \Omega \\ \Delta f_1 = 0\,\, \text{in}\,\, M-\bar{\Omega}
 \end{gather}

$F$, $G$ and $H$ are analytic functions in their arguments and the
symbol $C^2_t$ means that $f_1$ is $C^1$ on the boundary and its
tangential second derivatives $f_{1,ij}t^j$ are continuous for any
vector field $t=t^i\pd_i$ tangent to the boundary. Recall that $c$
is a constant not a priori prescribed. Note that this set of
equations is a generalization of the equations modelling Newtonian
and relativistic fluids.

In section \ref{main} we review some important theorems obtained
in reference \cite{PP-S} and prove new results regarding the
geometric properties of the equilibrium shapes, in particular that
the level sets of the function $f_1$ are isoparametric
submanifolds. In section \ref{properties} we obtain some
sufficient conditions for the existence of equilibrium shapes on
certain spaces and give some examples of manifolds for which
isoparametric functions (and therefore equilibrium shapes) do not
exist. Finally in section \ref{isom} the relationship between the
equilibrium shapes and the Killing vector fields of $M$ is
studied. These results are related to some theorems recently
proved by J. Szenthe \cite{Sz}. We also point out an interesting
property regarding isoperimetric domains. All these results are
new.

\section{Some geometric properties of the equilibrium shapes}\label{main}

Since $\pd \Omega$ is a level set of $f_1$ the approach that we
use to understand the equilibrium shapes consists in studying the
geometrical properties of the fibres $f_1^{-1}(c)$, $c\in f_1(M)$.
We assume that solutions to $(P1)$ exist and characterize the
structure of the level sets of these solutions. This is in strong
contrast with the classical approaches where the problem of
existence and uniqueness is firstly considered and then the
geometrical restrictions arise.

The set formed by the union of all the connected components of the
fibres of $f_1$ is called the partition of $M$ induced by $f_1$
and is denoted by $\beta_M(f_1)$. In general the dimension of the
leaves of the partition is not constant because there can exist
singular fibres ($\nabla f_1 = 0$) and therefore $\beta_M(f_1)$ is
a singular foliation.

In \cite{PP-S} it is proved that the partitions induced by $f_1$,
$f_2$ and $f_3$ agree fibrewise in $\Omega$. It is also proved
that $f_1$ is $C^{\omega}$ in $M-\pd\Omega$. As a consequence of
this property the singular set of $f_1$ has null-Lebesgue measure
and is topologically closed in $M$.

The partition $\beta_M(f_1)$ is not analytic across the
free-boundary $\pd \Omega$ but it has the remarkable property of
being analytically representable, that is, there exists a function
$I:M\to \RR$ analytic in the whole $M$ such that
$\beta_M(f_1)=\beta_M(I)$. The idea of the proof, which appears in
\cite{PP-S}, is that the interior symmetries propagate across the
free-boundary and remain symmetries of the exterior solution. Then
the partition is reconstructed from these symmetries in such a way
that it turns out to be analytic. Note that it is not immediate
that a partition which is analytic except for a fibre must be
analytically representable, in fact many examples are known
\cite{PP-S} in which the partition is not analytically
representable across the pathological fibre.

The main theorem in \cite{PP-S}, which provides a complete
geometrical characterization of the leaves of $\beta_M(f_1)$, is
the following:

\begin{theorem} \label{T1}
If $f_1$ is a solution of the problem $(P1)$ then $\beta_M(f_1)$
is an equilibrium partition
\end{theorem}

The concept of equilibrium partition is introduced in \cite{PP-S}.
We say that the analytic function $I:M \to \RR$ is of equilibrium
if $\Delta I$, $(\nabla I)^2$ and $I$ agree fibrewise in $M$. The
partitions induced by equilibrium functions are called equilibrium
partitions. Although this definition involves a particular
function $I$ the concept of equilibrium partition is mainly
geometrical, as the following proposition shows.

\begin{proposition}\label{inv}
Any analytic function representing an equilibrium partition
$\Sigma$ is an equilibrium function.
\end{proposition}
\begin{proof}
By definition there exists an equilibrium function $I$
representing $\Sigma$. The lemma follows if we manage to prove
that any analytic function $\hat I$ representing $\Sigma$ is of
equilibrium. Suppose that $\nabla I$ does not vanish in certain
open subset $U$ (it is always possible by the analyticity of $I$)
and consider a local coordinate system $(x^1,\ldots,x^n)$. Assume,
without loss of generality, that $I_{x^1}\neq 0$ in $U$, the
subscript denoting partial differentiation. Then the implicit
function theorem guarantees the following step:
$x^1=I^{-1}(I,x^2,\ldots,x^n) \Longrightarrow \hat I = \hat
I(I^{-1}(I,x^2,\ldots,x^n),x^2,\ldots,x^n)\equiv
F(I,x^2,\ldots,x^n)$. It is easy to check that
$F_{x^2}=\ldots=F_{x^n}=0$. One only has to take into account the
implicit function theorem and that $I, \hat I$ agree fibrewise.
Hence in $U$ we get that $\hat I = F(I)$, where $F$ is an analytic
function of its argument. Since $I$ is an equilibrium function we
have that locally (by the same argument involving the implicit
function theorem) $(\nabla I)^2$ and $\Delta I$ are functions of
$I$. Now a straightforward computation yields that $(\nabla \hat
I)^2=F'(I)^2(\nabla I)^2$ and $\Delta \hat I=F''(I)(\nabla
I)^2+F'(I)\Delta I$. This implies that locally $(\nabla \hat I)^2$
and $\Delta \hat I$ are functions of $\hat I$, and therefore $\hat
I$ is a local equilibrium function. The globalization of this
property follows from an analytical continuation result for
analytic partitions, that is, if $f$ and $g$ are two analytic
functions on $M$ which agree fibrewise in certain open subset $U$
then $\beta_M(f)=\beta_M(g)$. Indeed since $\beta_U(f)=\beta_U(g)$
we have that $\text{rank}(df,dg)\leq 1$ in $U$. Since $U$ is an
open set and $f, g$ are analytic functions we have that this
inequality is satisfied in the whole $M$, $\text{rank}(df,dg)\leq
1$ in $M$. This condition implies \cite{Ne} that $f$ and $g$ are
functionally dependent and so there exists an analytic function
$Q:\RR^2 \to \RR$ such that $Q(f,g)=0$ in $M$, which implies that
the partitions of $f$ and $g$ agree. Application of this result to
the triple $(\hat I, (\nabla \hat I)^2, \Delta \hat I)$ yields
that $\hat I$ is an equilibrium function.
\end{proof}

As a consequence of theorem \ref{T1} the problem of classifying
the partitions induced by the solutions of $(P1)$ is substituted
by the problem of classifying the equilibrium partitions on
different spaces. We have reduced the original problem involving a
difficult system of PDE to a purely geometrical problem. In the
next section we will show examples of manifolds which do not admit
equilibrium functions.Thus in these manifolds we geometrically
obtain an existence result: $(P1)$ cannot have solutions.

The following theorem characterizes the general properties that
all the equilibrium partitions must possess on any Riemannian
manifold \cite{PP-S}.
\begin{theorem} \label{char}
The partition induced by any equilibrium function $I$ on $M$ has a
trivial fibre bundle local structure, each leaf has constant mean
curvature and locally the leaves are geodesically parallel
\end{theorem}

By the term trivial fibre bundle local structure I mean that $M$
is divided into a countable number of disconnected,
$I$-partitioned (formed by level sets of $I$) open regions $M_i$
such that $M=\bigcup_i M_i\cup C(I)$, $C(I)$ standing for the
critical set of $I$ (nowhere dense in $M$), and the equilibrium
partition in each $M_i$ is a trivial fibre bundle. Note that each
$M_i$ may be made up by several connected components $M_i^j$. The
interested reader can have a look at reference \cite{PP-S} where
other geometrical properties of the equilibrium partitions are
obtained in constant curvature, conformally flat and locally
symmetric spaces. Note that the geodesical parallelism of
equilibrium partitions implies that they are Riemannian (singular)
foliations \cite{To, Mo} , a property which will be important in
the last section. In fact theorem \ref{char} can be locally
expressed as an equivalence.

\begin{proposition} \label{Riem}
A Riemannian codimension 1 (singular) foliation whose non-singular
leaves have constant mean curvature is locally an equilibrium
partition.
\end{proposition}
\begin{proof}
Consider an open subset $U\subset M$ small enough so that the
foliation in $U$ (which we assume regular) can be represented by a
function $f$. If $M$ is compact, simply connected and the
foliation has trivial holonomy then $f$ is defined on the whole
$M$ \cite{HeHi}. If we assume that the foliation is locally
trivial (this is the case when $M$ is compact and there are not
dense leaves \cite{To, Mo}) then we can extend $U$ to a set
$\Lambda$ formed by leaves of the foliation in such a way that the
first integral $f$ is well defined in the whole $\Lambda$ (in
general it will be defined on any globally trivial foliated set
$\Lambda$, note that in this case the holonomy is trivial). Since
the leaves are parallel then $f$ must satisfy that $(\nabla
f)^2=F(f)$ in $U$. Indeed if $g$ is the metric tensor and $D$ its
associated covariant derivative it is immediate that $X((\nabla
f)^2)=2g(D_X\nabla f,\nabla f)=2g(D_{\nabla f}\nabla f,X)$ for any
vector field $X$ on $U$. Since the foliation is geodesically
parallel then the gradient lines of $f$ are tangent to geodesics,
that is $D_{\nabla f}\nabla f=\lambda \nabla f$ for certain
real-valued function $\lambda$ on $U$. Identifying we get that
$X((\nabla f)^2)=2\lambda X(f)$ and therefore the symmetries of
$f$ are also symmetries of $(\nabla f)^2$ which implies, via
Frobenius theorem, that $(\nabla f)^2$ is a function of $f$. The
constancy of the mean curvature $H$ on the leaves is expressed as
$H=\text{div}(\frac{\nabla f}{\|\nabla f\|})=G(f)$. After some
computations, and taking into account that $(\nabla f)^2=F(f)$,
one readily gets that $\Delta f$ is also a function of $f$ in $U$
and hence we obtain the (local) equilibrium property.
\end{proof}

Now we prove a result which relates the equilibrium condition to
the well known isoparametric condition. Recall that a smooth
function $f:M \to \RR$ is called isoparametric if $(\nabla
f)^2=F(f)$ and $\Delta f=G(f)$ in $M$, $F$ and $G$ smooth
functions of their argument. This concept was firstly introduced
by Cartan \cite{Ca1,Ca2} and Segre \cite{Se} in a purely
geometrical context. Two good surveys on this topic are the works
of Nomizu \cite{No} and Thorgbersson \cite{Th}.

\begin{theorem}\label{isop}
An equilibrium function is locally isoparametric.
\end{theorem}
\begin{proof}
Assume that the equilibrium function $f$ has $N$ different
critical values (since $f$ is analytic the set of critical values
is discrete in $\RR$) and that $f(M)=(-\infty,+\infty)$. Otherwise
the technique can be adapted without problem. In this case
$M=\bigcup_{i=1}^{N+1}M_i\cup C(f)$, $C(f)$ standing for the
critical set of $f$, and possibly $M_i$ being made up by several
connected components $M_i^j$. In the open regions $M_i^j$ the
equilibrium function is submersive and the partition is globally
trivial, so we can take an open subset $\Lambda \subset M_i^j$
which is $f$-partitioned. An adaptation of an argument applied in
the proof of proposition \ref{inv} yields that $(\nabla f)^2$ and
$\Delta f$ are functions of $f$ in certain open subset of
$\Lambda$. The globalization of this property to the whole
$\Lambda$ stems from the existence of a local transversal (and
analytic) curve to the fibres of $f$, which is a consequence of
the triviality of the partition. In fact the isoparametric
condition holds in each region $M_i^j$ because triviality implies
the existence of a global transversal curve and therefore $f$,
which is submersive, can be adapted to a global coordinate system
in $M_i^j$ \cite{Ga}.
\end{proof}

The following example illustrates the fact that the isoparametric
character of an equilibrium partition is in general only local.

\begin{example}
The analytic function $f(x,y)=\cos \sqrt{x^2+y^2}$ in
$(\RR^2,\delta)$ is of equilibrium type: the partition is formed
by concentric circles and hence by proposition \ref{inv} it must
be an equilibrium function. On the contrary it is not a global
isoparametric function because $(\nabla f)^2=1-f^2$ but $\Delta f$
cannot be globally expressed as a function of $f$ due to the
existence of the critical fibres $r=i\pi$, $i\in \NN\cup\{0\}$.
Anyway, as proved in the theorem, $\Delta f$ is a well defined
function of $f$ in the domains $M_i=\{ i\pi<r<(i+1)\pi\}$ and a
straightforward computation yields $\Delta
f=-f-\frac{(-1)^i\sqrt{1-f^2}}{i\pi+\arccos((-1)^if)}$.
\end{example}

The most remarkable feature of theorem \ref{isop} is that the idea
of isoparametric submanifold, which was introduced in differential
geometry many decades ago, naturally arises in a physical context.
It is important to note that other authors have also employed the
isoparametric condition in order to study the partitions induced
by the solutions of certain PDE \cite{Ser, Sa, AlMa}, but the
techniques that we use are completely different to these authors'.

\section{The existence problem of the equilibrium
shapes}\label{properties}

In general it is a difficult task to know whether an equilibrium
(or isoparametric) function exists on a given Riemannian manifold.
This problem is not only interesting from the mathematical
viewpoint but also from the physical one. If certain space does
not admit equilibrium partitions then a self-gravitating fluid
will never reach static equilibrium, a non-existential result for
the set $(P1)$ of PDE induced by the geometrical/topological
properties of the manifold. It would be desirable to classify, in
certain well defined sense, spaces admitting equilibrium or
isoparametric functions, which would be the suitable spaces for
doing relevant physics. This restriction of physically admissible
manifolds reminds us of the constraints imposed by general
relativity: the geometry must be coupled with the matter. In fact,
as we will show in this section, many of the spaces for which we
manage to prove the existence of equilibrium functions possess
geometric structures linked to the partitions, and without this
link no isoparametric functions exist. It is surprising that this
property arises without taking into account the coupling of
general relativity but only the existence of equilibrium. As a
remarkable consequence a Newtonian fluid, which a priori does not
impose any constraint on the geometry, will not exist on certain
spaces and the existence will be likely related to the coupling of
certain geometric structures of $(M,g)$ with the equipotential
hypersurfaces, as happens in the Einstenian theory. All the
restrictions obtained in this section are of geometrical type and
it remains open to ascertain whether topological restrictions also
exist.

\begin{proposition} \label{foc}
An equilibrium partition with just one focal point (or caustic) $P
\in M$ exists on $(M,g)$ if and only if
$\text{det}(g)=A(r)^2B(\theta)^2$, where $\text{det}(g)$ is
expressed in polar Riemann coordinates $(r,\theta)$ around $p$. In
this case the equilibrium partition is locally formed by geodesic
spheres.
\end{proposition}
\begin{proof}
Recall that in polar Riemann coordinates centered at $P\in M$ the
metric tensor is expressed, in certain local neighborhood $U$, as
$\dd s^2=\dd r^2+G_{ij}(r,\theta)\dd \theta^i\dd \theta^j$. The
sufficiency condition stems from the fact that the function
$f=\frac{1}{2}r^2$ is of equilibrium. Indeed $(\nabla f)^2=r^2=2f$
and $\Delta f=\frac{\pd_r(rA(r))}{A(r)}$ which is an analytic
function of $r$ because $A(r)=r^{n-1}+O(r^n)$. Therefore $r^2$
induces a local equilibrium partition (the geodesic spheres) whose
focal point is $P$. Since polar Riemann coordinates are just local
then this partition could not be globally defined except when the
exponential map defines a global diffeomorphism from $\RR^n$ onto
$M$. This happens, for example, if the space is simply connected
and the sectional curvature is non-positive (Cartan-Hadamard's
theorem). Conversely if one has an equilibrium partition with a
caustic formed by the point $P$ then the geodesical parallelism of
the leaves implies that the partition must be formed by geodesic
spheres centered at $P$. This stems from the fact that the focal
varieties of Riemannian (singular) submersions are smooth
submanifolds of $M$ and the regular leaves of the partition are
tubes (constant distance) over either of the focal varieties
\cite{Wa}. On account of proposition \ref{inv} the function
$f=\frac{1}{2}r^2$ representing the same partition must be of
equilibrium. The condition of $\Delta f$ being a function of $r$
is expressed as $\pd_r \ln(r\sqrt{\text{det}(g)})=F(r)$ and a
straightforward integration yields that $\text{det}(g)$ must
factorize in two functions of $r$ and $\theta$.
\end{proof}

An important example in which $\text{det}(g)$ factorizes like in
proposition \ref{foc} is when the manifold is rotationally
symmetric around $P$ and therefore
$\text{det}(g)=A(r)^2\Omega(\theta)^2$, where $\Omega(\theta)^2$
is the determinant of the metric of the Euclidean sphere
$S^{n-1}\subset \RR^n$ in its usual coordinates. Any constant
curvature space satisfies this condition (with respect to any
point) so the local equilibrium partitions with $P$ as focal point
are the geodesic spheres centered at $P$. For the canonical
constant curvature manifolds $S^n, \RR^n$ and $\HH^n$ the local
partitions can be globalized (note that in $S^n$ there will appear
a second focal point).

Another consequence is that spaces whose metric does not satisfy
the assumption of factorization will not have (local) equilibrium
partitions with just one focal point. This property is important
from the physical viewpoint because a mass of fluid in general
will enclose a contractible domain (for example think of a fluid-
composed star) and hence in these spaces the fluid will not be
able to reach static equilibrium.

The following proposition establishes a stronger result, the
non-existence of equilibrium partitions, whatever the focal sets
be, for certain 2-dimensional manifolds.

\begin{proposition} \label{conf}
Let $(\RR^2,g)$ be a conformally flat 2-dimensional space. Then
equilibrium partitions exist if and only if the conformal factor
is an equilibrium function of $(\RR^2,\delta)$ and they are the
same as in the Euclidean case.
\end{proposition}
\begin{proof}
The conformally flat metric takes the canonical form
$g=\exp(2\phi)\delta$. A straightforward computation yields that
$(\nabla f)^2=\exp(-2\phi)(\nabla_Ef)_E^2$ and $\Delta f =
\exp(-2\phi)\Delta_E f$, where the subscript $E$ means that the
corresponding operation must be carried out in the Euclidean
space. Locally the equilibrium function $f$ is isoparametric so
$\exp(-2\phi)(\nabla_Ef)_E^2=F(f)$ and $\exp(-2\phi)\Delta_E
f=G(f)$. Dividing both equations we get that $\frac{\Delta_E
f}{(\nabla_Ef)_E^2} = \frac{G(f)}{F(f)}$. This equation implies
that the partition induced by $f$ is the same as the partition
induced by certain harmonic function in $(\RR^2,\delta)$. Indeed
if $\Delta_E u = 0$ and we re-parametrize $\beta_{\RR^2}(u)$ by
another function $f$ such that $u=T(f)$ then $f$ will verify that
$\frac{\Delta_E f}{(\nabla_Ef)_E^2} = -\frac{T''(f)}{T'(f)}$.
Identifying we conclude that the re-parametrization is given by
$-\frac{\dd}{\dd f} \ln T'(f)=\frac{G(f)}{F(f)}$. The condition of
inducing partitions given by harmonic functions is invariant under
the change $f=R(h)$, for any $R$, and hence $\frac{\Delta_E
h}{(\nabla_E h)_E^2} = M(h)$. Now the isoparametric condition for
$f$ implies that $R''(h)(\nabla_E h)_E^2 + R'(h)\Delta_E h = \hat
G (h) \exp{2\phi}$. Both equations are compatible if and only
$\phi$ is a function of $h$ and therefore the partitions will be
isoparametric also in the Euclidean plane. The globalization is
immediate for analytic functions.
\end{proof}

The implication of this result is remarkable, in 2-dimensional
conformally flat spaces there not exist solutions to problem
$(P1)$ unless the conformal factor be an Euclidean isoparametric
function. In this case the equilibrium partitions are those of the
conformal factor and are given, up to rigid motions, by
$\beta_{\RR^2}(x^2+y^2)$ and $\beta_{\RR^2}(x)$. We have obtained
this coupling condition without taking into account the Einstenian
coupling between matter and geometry, in particular it holds for
Newtonian fluids. In fact the assumption that the conformal factor
$\phi$ is a function of the gravitatory potential is very common
in the physics literature (see for example \cite{Lind2}). We have
hence justified this hypothesis in the 2-dimensional case: it is a
geometrical constraint in order that equilibrium solutions exist.
Propositions \ref{foc} and \ref{conf} show that physically
relevant manifolds will have to satisfy hard constraints.

For instance in the Riemannian manifold $(\RR^2,\exp(y-x^2)(\dd
x^2+ \dd y^2))$ there not exist equilibrium partitions and
therefore a self-gravitating fluid would never reach the static
equilibrium, fluid-composed stars would not be possible. It is
remarkable to note that this space does not possess (global)
Killing vector fields, a fact which is very related to the
existence of equilibrium functions, as we will show.

\section{Equilibrium shapes, isoperimetric domains and
isometries}\label{isom}

The results of the preceding section suggest that equilibrium
partitions are always linked to certain geometric structures of
the manifold. If these geometric structures fail to exist then
equilibrium partitions do not exist. For example consider the
Euclidean space $(\RR^n,\delta)$. It can be proved \cite{PP-S}
that the equilibrium partitions of this manifold are geometrically
trivial in the sense that all the leaves are globally isometric to
$bS^p\times\RR^q$, $p$ and $q$ fixed natural numbers such that
$p+q=n-1$, $b\geq 0$. These partitions have the remarkable
property of being generated by isometries of $(\RR^n,\delta)$. In
general, as a consequence of proposition \ref{foc}, the
equilibrium partitions with $P$ as focal point are induced by
isometric group actions on $M$ whenever the space is rotationally
symmetric around $P$. This fact suggests that the isometries of
the manifold are somehow related to the equilibrium partitions.
The following proposition establishes the equivalence between both
concepts for 2-dimensional manifolds.

\begin{proposition} \label{Kill}
Let $(M,g)$ be a 2-dimensional Riemannian space. Then the
equilibrium partitions are 1-dimensional (singular) foliations
generated by Killing vector fields of $(M,g)$.
\end{proposition}
\begin{proof}
The equilibrium partitions of $M$ are formed by leaves of
dimension 1, except for a nowhere dense and closed subset of
singular points. This defines a 1-dimensional (singular)
foliation. In dimension 1 mean curvature and Gauss curvature agree
and therefore the orbits possess constant Gauss curvature. As a
consequence of Gauss theorem we have that the sectional curvature
of $(M,g)$ restricted to each leaf is also constant (the intrinsic
sectional curvature of a 1-dimensional manifold is always
trivial). This implies that the fibres are transitivity lines of
an isometric group action on the space and therefore correspond to
the orbits of a Killing vector field.
\end{proof}

This proposition proves analogous results appearing in proposition
2 and theorem 2, reference \cite{Sz}, in the Riemannian case and
without assuming that $M$ is simply connected. Indeed if $f$ is a
submersive equilibrium function you can adapt it to a (local)
coordinate system $(f,g)$. Since $\pd_g$, tangent to the levels of
$f$, is a Killing then the metric can be locally expressed as a
warped product, $ds^2=A(f)(df^2+B(f)dg^2)$. In fact since $f$
induces a Riemannian (non-singular) foliation with a global
transversal curve $\Sigma$ (diffeomorphic to $S^1$ or $\RR$) then
$M$ can be globally expressed as a product $\{f=0\}\times \Sigma$,
which just leaves the possibilities $M\simeq \RR^2,\, \RR\times
S^1\, \text{and}\, S^1\times S^1$, and therefore the warped
product expression globalizes.

The converse of proposition \ref{Kill} holds in very general
situations, as we prove in the next theorem. Note that for
2-dimensional manifolds which do not admit Killing vector fields
equilibrium partitions do not exist. This is directly related to
proposition \ref{conf} and implies that the spaces
$(\RR^2,\exp({2\phi})\delta)$ for which $\phi$ is not an
equilibrium function of $(\RR^2,\delta)$ do not possess Killing
vector fields.

\begin{theorem} \label{Lie}
Let $\Xi=\{\xi_1,\ldots,\xi_p\}$, $p\geq n-1$, be a Lie algebra of
Killing vector fields of $(M,g)$. $\Xi$ satisfies that
$\text{rank}(\xi_1,\ldots,\xi_p)=n-1$ in $M$, up to a null measure
set, and it generates a closed subgroup of the group of
isometries. Then the (singular) foliation induced by $\Xi$ is an
equilibrium partition.
\end{theorem}
\begin{proof}
$\Xi$ generates an isometric group action $G$ on $M$. $G$ is
connected, simply connected (take the universal covering) and
closed in the group of isometries (by assumption). This defines a
proper group action on the manifold and therefore $M$ can be
divided into two components \cite{Pa}, the principal part
$M^{\ast}$, which is open and dense in $M$, and the singular part,
which is formed by totally geodesic submanifolds. $M^{\ast}$ is
foliated by codimension 1 closed submanifolds of $M$, in fact this
foliation is a Riemannian submersion from $M$ to $M/G$ \cite{Mo}.
Note that $M/G$ is a differential Hausdorff 1-manifold, and
therefore diffeomorphic to $\RR$ or $S^1$. The submersion is
analytic because we always assume in this paper that $(M,g)$ is
analytic, and therefore also the Killing vector fields. Call $f$
the function representing the foliation in $M^{\ast}$; since it is
Riemannian then $f$ will satisfy that $(\nabla f)^2=F(f)$, as
proved in proposition \ref{Riem}. Since the foliation is globally
trivial (there exists a global transversal curve and the leaves
are all diffeomorphic among them) this condition holds in the
whole $M^{\ast}$. Since the action of $G$ is transitive on each
leaf (the leaves are extrinsically homogeneous, that is
homogeneous by isometries of the ambient space) then the mean
curvature must be constant at all point of the leaf \cite{Pe}.
This follows from the fact that the second fundamental forms at
two different points connected by an isometry correspond through
this isometry. In terms of $f$ this condition is expressed as
$H=\text{div}(\frac{\nabla f}{\|\nabla f\|})=H(f)$. The following
computation $\text{div}(\frac{\nabla f}{\|\nabla
f\|})=\frac{\Delta f}{\|\nabla f\|}-\frac{\nabla f \nabla(\|\nabla
f\|)}{(\nabla f)^2}$ readily implies that $\Delta f = G(f)$. Since
the non-principal set is nowhere dense the isoparametric condition
extends to the whole $M$ and therefore the foliation is of
equilibrium. Note that the extended $f$ could fail to be analytic
in the singular set.
\end{proof}

\begin{remark}
In general it is necessary to require that $G$ be closed in the
group of isometries \cite{Pe}. For example, take the flat 2-torus
$S^1\times S^1$ and consider the action by the real line which is
given by an irrational translation. This induces a Killing vector
field, 1-dimensional, but the group generated is not closed in the
isometry group of the torus, which we know is compact (it is
$O(2)\times O(2)$). In fact this action is not proper since orbits
are not embedded. Similar examples can be constructed in greater
dimension.
\end{remark}

Note that theorem \ref{Lie} generalizes theorem 1 in \cite{Sz} to
arbitrary dimension in the Riemannian setting. In general the
converse of this theorem is true only for 2-dimensional manifolds
(proposition \ref{Kill}). Indeed consider a manifold which is not
rotationally symmetric with respect to the point $P$ but the
determinant of the metric in polar Riemann coordinates factorizes
as in proposition \ref{foc}. Then the geodesic spheres around $P$
are equilibrium submanifolds but they are not induced by an
isometric group action. This shows that in general the converse
theorem does not hold. It would be interesting to find conditions
in order that the equilibrium partitions of a manifold be
(singular) foliations induced by isometric group actions.

All these results show the deep relationship between isometries
and equilibrium and suggest that physically relevant spaces should
possess enough Killing vector fields. Consequently an effective
procedure in order to obtain equilibrium partitions, and hence
equilibrium configurations of self-gravitating fluids, is to
compute the Killing vector fields of the space. It is likely that
spaces which just admit a few isometries (or even no one) do not
admit equilibrium functions either, lacking static configurations.
Let us illustrate now theorem \ref{Lie} with an example.

\begin{example}
Consider the space $\HH^2\times \RR$ endowed with the metric $\dd
s^2=\frac{\dd x^2+\dd y^2}{F^2}+\dd z^2$, where
$F=\frac{2-x^2-y^2}{2}$ and $\HH^2=\{(x,y)\in \RR^2 :x^2+y^2<2\}$.
The Killing vector fields of this manifold are \cite{Mon}:
$X_1=(F+y^2)\pd_x-xy\pd_y$, $X_2=-xy\pd_x+(F+x^2)\pd_y$,
$X_3=-y\pd_x+x\pd_y$ and $X_4=\pd_z$. A straightforward
computation yields that $(\nabla f)^2=F^2(f_x^2+f_y^2)+f_z^2$ and
$\Delta f=F^2(f_{xx}+f_{yy})+f_{zz}$. Some easy, although long,
computations show that the codimension 1 partitions (up to null
measure set) induced by the Killings vector fields are:
\begin{itemize}
\item $\{X_3,X_4\}$ $\Longrightarrow$ $f=x^2+y^2$, which is an
equilibrium function.\item $\{X_i,X_j\}$, $i\neq j=1,2,3$
$\Longrightarrow$ $f=z$, which is an equilibrium function. \item
$\{X_1,X_4\}$ $\Longrightarrow$ $f=\frac{x^2+y^2-2}{y}$, which is
an equilibrium function (with a singular set). \item $\{X_2,X_4\}$
$\Longrightarrow$ $f=\frac{x^2+y^2-2}{x}$, which is an equilibrium
function (with a singular set).
\end{itemize}
\end{example}

From the physical viewpoint it is reasonable to compare the shapes
of a compact self-gravitating fluid with the isoperimetric
domains. By the term isoperimetric I mean the sets which minimize
the area for variations which leave fixed the volume. In the
Euclidean space the only compact equilibrium submanifold is the
round sphere, which is exactly the solution to the isoperimetric
problem. The physical meaning is clear: fluid-composed stars would
minimize their surfaces in order to achieve equilibrium.
Regretfully for general Riemannian manifolds an equilibrium
submanifold does not solve the isoperimetric problem. The most
general result that can be proved is the following.

\begin{proposition}
Let $S$ be a compact equilibrium codimension 1 submanifold. Then
$S$ is a critical point of the $(n-1)$-area $A(t)$ for all
variations $S_t$ that leave constant the $n$-volume $V(t)$
enclosed by $S$.
\end{proposition}
\begin{proof}
$S$ is the level set of an analytic function and therefore it has
no boundary. Since it is compact it encloses a finite volume. The
equilibrium condition implies that the mean curvature is a
constant $H$. Let $S_t$, $t \in (-\epsilon,\epsilon)$ and $S_0=S$,
be a variation of $S$. The first variation of the area at $t=0$ is
given by \cite{BCE} $A'(0)=-(n-1)H\int_S f\dd S$, where $f$ is the
normal component of the variation vector of $S_t$ and $\dd S$ is
the $(n-1)$-area element of $S$. Since the variation is volume
preserving then $V'(0)=\int_S f \dd S =0$ and therefore we get
that $A'(0)=0$.
\end{proof}

This result cannot be improved in general. We can find manifolds
for which equilibrium shapes are minimizers of the area and other
manifolds for which they are maximizers or saddle points. Even the
weaker condition of being stable, that is $A''(0)\geq 0$, is not
generally verified. It would be interesting to classify all the
spaces whose compact equilibrium submanifolds are stable. The
following list gives some of them:

\begin{itemize}
\item Constant curvature simply connected manifolds. The geodesic
spheres are the only stable submanifolds \cite{BCE}. They are also
of equilibrium on account of proposition \ref{foc}. \item
Rotationally symmetric planes with decreasing curvature from the
origin. The geodesic circles are stable and enclose isoparametric
domains \cite{Ri}, they are also of equilibrium. \item
Rotationally symmetric spheres with curvature increasing from the
equator and equatorial symmetry. The geodesic circles are stable
and enclose isoperimetric domains \cite{Ri}, they are also of
equilibrium. \item Rotationally symmetric cylinders with
decreasing curvature from one end and finite area. The circles of
revolution are stable, enclose isoperimetric domains \cite{Ri} and
a straightforward computation yields that they are also of
equilibrium.
\end{itemize}

It is not difficult to construct examples of manifolds with
equilibrium partitions whose leaves are not stable. For instance
consider the plane with the following metric tensor in polar
coordinates $\dd s^2=dr^2+r^2(1+r^2)^2\dd \theta^2$. The function
$I=\frac{1}{2}r^2$ is of equilibrium, it induces the equilibrium
partition given by the geodesic circles. Now, if you set
$f(r)=r(1+r^2)$, the expression $f'^2-ff''=1+3r^4$ is greater than
$1$ when $r>0$. This implies \cite{Ri} that no stable curves
exist. Other similar examples in dimension 2 can be found in the
work of Ritore. Other interesting example is given by the
symmetric spaces of rank 1. The geodesic spheres are transitivity
hypersurfaces of the group of isometries and therefore they are
equilibrium submanifolds (theorem \ref{Lie}). However not all the
geodesic spheres are stable \cite{BCE}.

The most remarkable physical conclusion of this work is that if we
want to recover the physical intuition that we have in the
Euclidean space, such as the existence of contractible fluid
domains in equilibrium, the coupling between symmetries of the
space and the shapes of the fluid or the stability of the fluid
regions we will have to restrict the base manifold. It would be no
surprising that only a few base spaces were able to give
physically relevant results.

\section{Acknowledgements}

I am very grateful to Rolando Magnanini, for providing me with the
literature on the isoparametric condition, to Stefano Montaldo and
Renato Pedrosa for their useful comments regarding isometric group
actions, to Cesar Rosales, for providing me with references
\cite{BCE} and \cite{Ri} and his useful comments regarding the
isoperimetric problem, and to Janos Szenthe, for providing me with
reference \cite{Sz}. The author is supported by an FPU grant from
MECD (Spain).

\end{document}